Regular Article - Experimental Physics

# The environmental low-frequency background for macro-calorimeters at the millikelvin scale

L. Aragão[1,2], A. Armigliato[2], R. Brancaccio[2,3], C. Brofferio[4,5], S. Castellaro[2], A. D'Addabbo[6], G. De Luca[7], F. Del Corso[3], S. Di Sabatino[2], R. Liu[8], L. Marini[6], I. Nutini[4,5], S. Quitadamo[6,9,a] , P. Ruggieri[2], K. J. Vetter[10,11], M. Zavatarelli[2], S. Zucchelli[3,12]

[1] Climate Simulations and Predictions Division, Euro-Mediterranean Centre on Climate Change (CMCC) Foundation, 40127 Bologna, Italy
[2] Dipartimento di Fisica e Astronomia, Alma Mater Studiorum-Università di Bologna, 40126 Bologna, Italy
[3] Istituto Nazionale di Fisica Nucleare-Sezione di Bologna, 40127 Bologna, Italy
[4] Istituto Nazionale di Fisica Nucleare-Sezione di Milano Bicocca, 20126 Milan, Italy
[5] Dipartimento di Fisica, Università di Milano-Bicocca, 20126 Milan, Italy
[6] Istituto Nazionale di Fisica Nucleare-Laboratori Nazionali del Gran Sasso, 67100 L'Aquila, Italy
[7] Istituto Nazionale di Geofisica e Vulcanologia-Osservatorio Nazionale Terremoti-Sede di L'Aquila, 67100 L'Aquila, Italy
[8] Wright Laboratory, Department of Physics, Yale University, New Haven, CT 06520, USA
[9] Gran Sasso Science Institute, 67100 L'Aquila, Italy
[10] Department of Physics, University of California, Berkeley, CA 94720, USA
[11] Nuclear Science Division, Lawrence Berkeley National Laboratory, Berkeley, CA 94720, USA
[12] Alma Mater Studiorum-Università di Bologna, 40126 Bologna, Italy

**Abstract** Many of the most sensitive physics experiments searching for rare events, like neutrinoless double beta ($0\nu\beta\beta$) decay, coherent elastic neutrino nucleus scattering and dark matter interactions, rely on cryogenic macro-calorimeters operating at the mK-scale. Located underground at the Gran Sasso National Laboratory (LNGS), in central Italy, CUORE (Cryogenic Underground Observatory for Rare Events) is one of the leading experiments for the search of $0\nu\beta\beta$ decay, implementing the low-temperature calorimetric technology. We present a novel multi-device analysis to correlate environmental phenomena with the low-frequency noise of low-temperature calorimeters. Indeed, the correlation of marine and seismic data with data from a couple of CUORE detectors indicates that cryogenic detectors are sensitive not only to intense vibrations generated by earthquakes, but also to the much fainter vibrations induced by marine microseisms in the Mediterranean Sea due to the motion of sea waves. Proving that cryogenic macro-calorimeters are sensitive to such environmental sources of noise opens the possibility of studying their impact on the detectors physics-case sensitivity. Moreover, this study could pave the road for technology developments dedicated to the mitigation of the noise induced by marine microseisms, from which the entire community of cryogenic calorimeters can benefit.

[a] e-mail: simone.quitadamo@gssi.it (corresponding author)

## 1 Introduction

Low-temperature calorimeters operated at the mK-scale are known to be extremely sensitive to vibrations. Indeed, an external force acting on the cryogenic system can induce a power dissipation on the hosted detectors. Depending on the nature of the source of vibration and on which elements are involved in the transmission of the associated power, different characteristic frequencies of the cryogenic setup can be excited and transmitted to the detectors. As a result, the operating temperature or the noise of the detectors can momentarily change. From the case study of a pair of low-temperature calorimeters from the CUORE experiment, we will show here that a correlation arises between the detectors low-frequency noise components and the marine microseismic activity.

We report a multi-device analysis technique for the study of the correlation of the low-frequency noise of cryogenic calorimeters with environmental sources of vibrations, namely the marine microseismic activity of the Mediterranean Sea. Section 2 is dedicated to a general description of low-temperature calorimeters, with a particular focus on the sources of noise which can affect their performance. Section 3 is dedicated to the description of the seismometers used for this study, their location at LNGS and their mutual compatibility. Section 4 focuses on the marine data provided by Copernicus Marine Environment Monitoring

Service and their compatibility with seismometers at LNGS. Finally, in Sect. 5 we describe in detail the analysis procedure we applied to correlate marine, seismic and calorimetric data in order to study the impact of marine microseisms in the Mediterranean Sea on the low-frequency noise of a couple of CUORE detectors.

## 2 Low-temperature calorimeters

Cryogenic detectors operated at the mK-scale represent cutting-edge technology in the field of low-temperature physics and precision measurements. They operate by detecting the phonons generated by an energy deposition in the detector, which results in a measurable temporary increase of its temperature. A thermal sensor is coupled to the energy absorber in order to measure such temperature variation and convert it into an electric signal. The size of the absorber depends on the physics applications: micro-calorimeters ($\sim$ µg-scale) are usually implemented for X-ray spectroscopy and $\beta$-decay studies, while macro-calorimeters ($\sim$g to kg-scale) are used for $\gamma$-ray spectroscopy, for coherent neutrino scattering experiments, and for $0\nu\beta\beta$ decay and dark matter searches [1]. Macro-calorimeters are often read out by means of semiconductor thermistors, with $\sim$ms response time. Therefore they are sensitive to thermalized phonons, thermal quanta which reached a new equilibrium after degrading their energy via several interactions in the absorber. In this case the temperature variation induced by an energy deposition is proportional to the deposited energy itself. Neutron Transmutation Doped germanium (Ge-NTD) thermistors are among the most widely used devices, since they can be operated in a wide range of temperatures by means of a simple read-out circuit at room temperature [2,3].

Operating detectors at the mK-scale ensures that the thermal noise is minimized, allowing for exceptional energy resolution and sensitivity. This makes such detectors particularly suited for a wide range of scientific endeavors, from the detection of rare events to astrophysical dark matter searches and neutrino detection through coherent elastic neutrino nucleus scattering. [4,5].

### 2.1 Noise and signal bandwidth of cryogenic detectors

At ultra-low temperatures many sources of noise can affect the performance of cryogenic detectors, and therefore noise mitigation is a critical aspect in their operation. A common source of noise for low-temperature calorimeters is electronic noise from the read-out chain, including the Johnson-Nyquist noise [6,7] due to the motion of charge carriers in the load resistor of the thermistor read-out circuit. Electronic noise can be mitigated by developing dedicated low-noise electronics [8].

Vibrational noise constitutes a noteworthy challenge in achieving high precision measurements. Mechanical vibrations from various sources, such as acoustic disturbances or microphonic effects, can perturb the detectors and degrade their performance. Therefore cryogenic detectors are often coupled to sophisticated suspension and isolation systems and to mechanical filters in order to attenuate vibrations [9].

Depending on the detector size and the type of read-out sensors, the signal bandwidth can span from less that 1 Hz up to hundreds of kHz. Various sources of noise, such as electronic and vibrational disturbances, can induce fluctuations in the measured signal at frequencies within the detector signal bandwidth, limiting its capability to accurately resolve the signal and evaluate the associated energy deposition.

### 2.2 The case study: CUORE detectors

CUORE is an experiment designed for the search of $0\nu\beta\beta$ decay of $^{130}$Te [10]. Hosted at LNGS, in central Italy (42.45° N, 13.58° E), it is the first ton-scale experiment based on cryogenic calorimeters. Taking data since 2017, CUORE consists of 988 TeO$_2$ crystals instrumented with Ge-NTDs and operated as low-temperature calorimeters at $\sim$10 mK by means of a custom designed cryostat with a $^3$He/$^4$He dilution refrigerator [9]. CUORE low-temperature calorimeters, each of them with a mass of 750 g, are organized in a modular structure of 19 identical towers (Fig. 1).

To maximise the performance in terms of sensitivity and energy resolution, it is mandatory for CUORE to minimize every source of noise induced on the detectors from the cryogenic infrastructure and the external environment. The CUORE cryostat is equipped with pulse tube (PT) cryocoolers, acting as a stage of the refrigeration system. Their operation generates vibrational noise at 1.4 Hz and harmonics. An active noise cancellation technique has been developed and implemented in order to stabilize and attenuate the PT-induced noise on the cryogenic detectors [11]. Suspension and decoupling systems are deployed in order to isolate as much as possible the detectors from the surrounding environment [9]. The detector payload at 10 mK is decoupled from the cryogenic system by a suspension system held by a support structure mounted on top of three mechanical insulators based on the negative-stiffness decouplers technology [12] manufactured by *Minus K Technology*. The entire structure of CUORE features a resonant behaviour at $\simeq$ 0.5 to 0.7 Hz, identified by means of seismometric measurements, which could enhance vibrations in the sub-Hz domain picked-up from the external environment.

The spectral shape of the noise of CUORE detectors is very complex (Fig. 2), being the sum of several contributions [3]. It is possible to identify the high-frequency cut-off at 120 Hz due to the anti-aliasing Bessel filters [8], and several peaks in the noise spectrum. As already mentioned, the 1.4 Hz peak and its harmonics are generated by the PT operation [11]. Other peaks are attributed to residual mechanical vibrations and oscillations of the suspension system and support structure (e.g. 0.6 Hz and 0.85 Hz) [13].

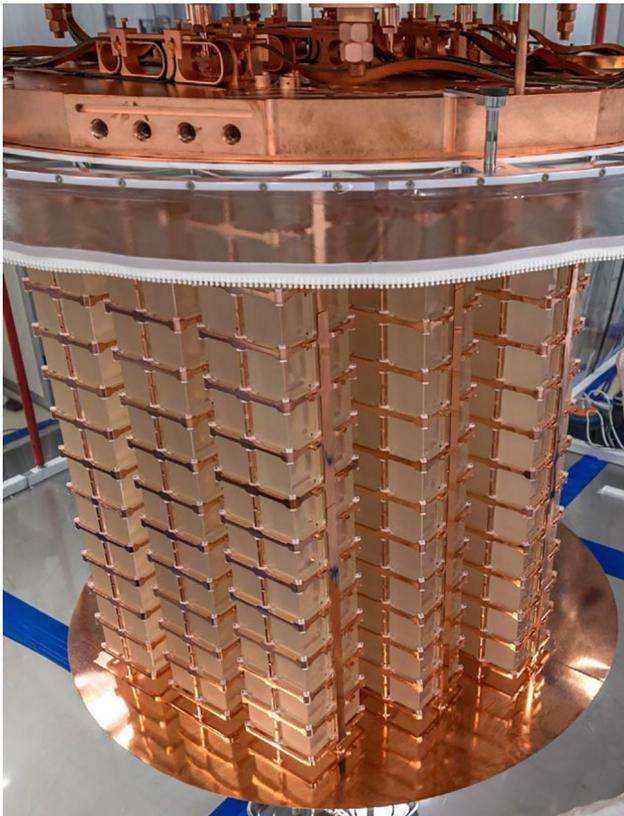

**Fig. 1** CUORE array of low-temperature calorimeters. Figure from [10]

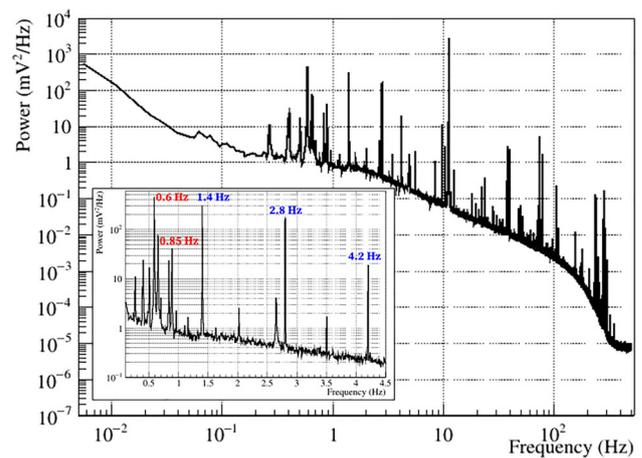

**Fig. 2** Average noise power spectrum (ANPS) of a CUORE low-temperature calorimeter. The insert shows a linear scale zoom of the ANPS in the [0,4.5] Hz region, which is the frequency range corresponding to the CUORE signal band. Noise peaks related to pulse tube cryocoolers harmonics are labeled in blue, while the ones related to the suspension system are labeled in red

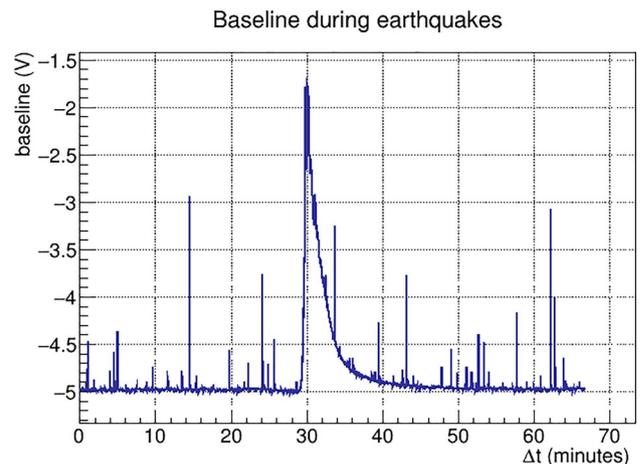

**Fig. 3** Baseline of a CUORE low-temperature calorimeter. The baseline is the voltage reading of the detector, associated to its temperature. The main discontinuity in the baseline is an effect of a violent earthquake originated in Mexico [14], while minor spikes are associated to less intense earthquakes

The support structure can also pick up vibrations generated by environmental phenomena. Indeed, many years of data taking established that CUORE detectors, and in general macro-calorimeters operated at the mK-scale, are sensitive to earthquakes-induced vibrations. As a reference, Fig. 3 shows the response of a CUORE low-temperature calorimeter to a violent earthquake (moment magnitude $M_w = 8.2$) originated on September 8th 2017 in the southern coast of Mexico [14], more than 10,000 km away from the detection point at LNGS. The induced seismic activity increased the detectors temperature by $\simeq 1$ mK, which then recovered to the working temperature on a time scale of $\simeq 15$ min. Minor spikes in Fig. 3 are associated to local, less intense earthquakes.

The motion of sea waves and the developments of marine storms are additional known sources of environmental vibrations [15–17]. Such events are the sources of marine microseismic activity, characterized by much fainter vibrations with respect to earthquakes and by characteristic frequencies in the sub-Hz domain. While the effects of seismic activity on cryogenic detectors is well established, the impact of microseismic activity is still not extensively studied.

## 3 Seismometric devices

Low-frequency environmental noise consists of ground vibrations caused by a variety of sources, such as seismic events (e.g. earthquakes) and microseismic events (marine activity), atmospheric perturbations and human activities. Seismometric devices can be deployed as complementary tools for highly-sensitive physics experiments, in order to study and mitigate environmental perturbations which can affect their experimental performance.

A seismometer is an instrument designed to detect seismic waves generated by various sources, including earthquakes, volcanic eruptions, movements of large masses of water, atmospheric phenomena and human activities. A seismometer is based on the principle of relative motion between a stationary mass and the moving Earth. The core of a seismometer is a mass-spring system, where a heavy mass is suspended to a fixed frame through a flexible spring. When seismic waves are detected, they are transmitted to the seismometer frame, which is therefore set in motion. The suspended mass, due to its high inertia, lags behind. The relative displacement between the mass and the frame is usually measured by means of an electromagnetic or optical sensor, carrying information about the main characteristics of the seismic waves (arrival time, amplitude, frequency...). Velocimeters are a type of seismometers sensitive to the relative velocity between frame and suspended mass.

### 3.1 Seismometers at LNGS

Two seismometers (called SEISMO 1, SEISMO 2) are installed in the experimental area of CUORE, mainly to identify and eliminate instabilities in the detectors data due to seismic and anthropic noises. Their sensitive element consists of a *SARA SS45* triaxial velocimeter [18] with a sampling rate of 400 Hz, characterized by a sensitivity bandwidth in the 0.2–400 Hz and by an almost flat response in the 1.0–50.0 Hz range.

Moreover the LNGS underground facility hosts a seismic station (called GIGS) from the Italian National Institute of Geophysics and Vulcanology (INGV) [19]. Previously hosted by the homonym GIGS (Geophysical Interferometer at Gran Sasso) experiment, GIGS is part of a network of stations spread all through the Italian Peninsula to monitor and study geological phenomena in the Mediterranean region.

GIGS [20, 21] is a *Nanometric Trillium-240 S* velocimeter with a sampling rate of 100 Hz, characterized by enhanced sensitivity at low frequencies and by a flat broadband response from 4 mHz to 35 Hz. The GIGS sensing elements are arranged in a symmetric triaxial configuration, ensuring the same response on vertical and horizontal directions.

### 3.2 Correlation of GIGS and SEISMO data

GIGS is located $\simeq 130$ m away from the CUORE experimental area, where SEISMO are installed. Therefore these seismometers allow to investigate the vibrational noise in different locations at LNGS and possibly identify if its sources are internal or external to the underground facility. While the horizontal axes of GIGS are oriented in the North–South and East–West directions, the orientation of SEISMO was performed in relation to their hosting experimental hall. Due to the difficulties in achieving a precise comparison between the orientation of the horizontal axes of GIGS and SEISMO, the vertical Z axes is the only one directly comparable among the three seismometers. For this reason, here we will focus only on their velocity measurements on the vertical axes, whose root mean square (RMS) is a good estimator of the variation of the vibrational activity at LNGS.

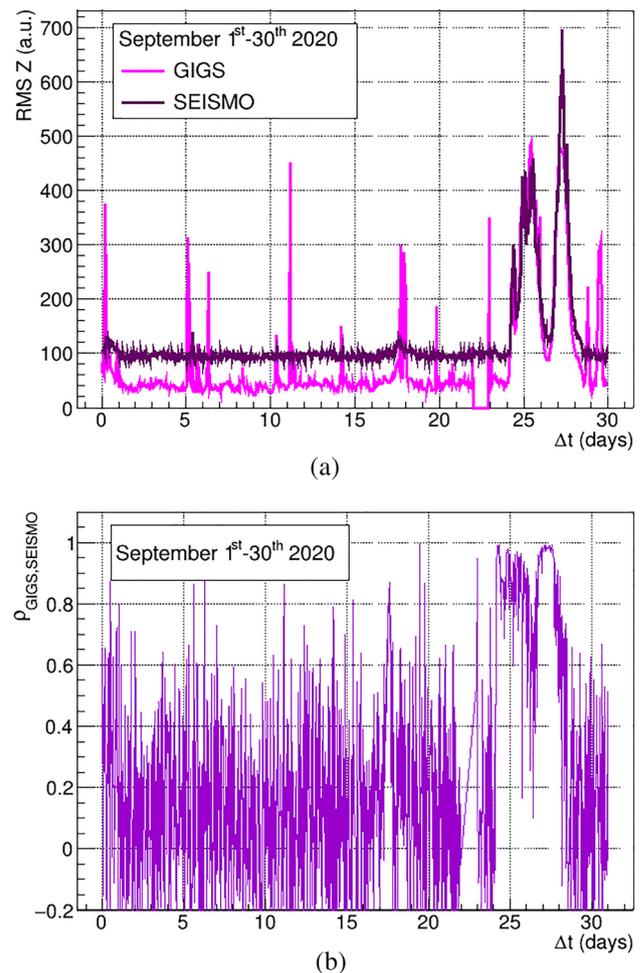

**Fig. 4** Time evolutions of GIGS and SEISMO RMS Z (**a**) and of their correlation coefficient (**b**) in September 2020. The spikes of high correlation can be attributed to earthquakes, since they affect all the seismometers

Figure 4a shows the comparison between the time profiles of RMS Z of GIGS and SEISMO during September 2020. Due to their relative distance, local disturbances, such as anthropogenic noise, generated around the area in which one seismometer is located are not detected by the other device. Differently, non-local events involving the entire LNGS are detected by all the seismic stations. Non-local perturbations can be attributed to different sources, like seismic or marine activities. By correlating seismometric data with marine data, we could identify the source of the prominent perturbation in the end of the month as a storm developing in the Mediterranean Sea (more details will be discussed in Sect. 4). During the storm development the responses of GIGS and SEISMO are strongly correlated (Fig. 4b) (in the period of maximum perturbation the Pearson correlation coefficient is $\rho_{GIGS,SEISMO} > 0.90$), highlighting the non-local nature of the environmental event responsible for such perturbation.

## 4 Copernicus marine environment monitoring service

The Copernicus Program [22] is the Earth monitoring component of the European Union space programme, which provides environmental data based on satellites and in-situ measurements, and on numerical models. The Copernicus Marine Environment Monitoring Service (CMEMS) [23] is the Copernicus sector dedicated to marine observation, providing free-of-charge, state-of-the-art ocean data at global and regional scales.

Here we refer to two CMEMS products, namely *Mediterranean Sea Waves Reanalysis and Forecast* [24] and *Mediterranean Sea Waves Analysis and Forecast* [25]), whose data originate from the implementation of the wave numerical model WAM [26] to the Mediterranean Sea (Med-WAM). The model domain covers the whole Mediterranean Sea with a regular grid with $\frac{1}{24}°$ horizontal resolution ($\simeq 4.6$ km). It is designed to enclose the Atlantic Ocean swells, obtained from CMEMS, propagating to the Mediterranean Sea through the Gibraltar strait. The atmospheric forcing is rendered by surface winds, obtained from the *fifth generation ReAnalysis* (ERA5) produced by the European Center for Medium-range Weather Forecast (ECMWF) [27].

### 4.1 Data selection

In this work we refer to two oceanographic fields among those provided, with a time granularity of 1 h, by CMEMS:

1. VHM0: the spectral significant wave height, approximately equal to the average height of the highest one-third of the recorded waves heights:

$$\text{VHM0} = \frac{1}{N/3} \sum_{n=1}^{N/3} H_n \quad (1)$$

   where $H_n$ is the wave height, sorted from the highest wave ($H_1$) to the lowest ($H_N$), and $N$ is the size of the set from which the highest one-third waves sub-set is extracted;
2. VTPK: the sea surface wave period at maximum variance spectral density; the inverse of VTPK defines the sea waves frequency $\nu_{sea}$.

Here we select and average CMEMS data in two rectangular regions in the Adriatic and Tyrrhenian Seas (labeled as *Domain 1* and *Domain 2* in Fig. 5):

1. Adriatic Sea: $[42.5, 43.0]°$ N $\times [14.0, 16.0]°$ E($\simeq 53 \times 222$ km$^2$);
2. Tyrrhenian Sea: $[40.0, 41.0]°$ N $\times [12.0, 14.0]°$ E($\simeq 108 \times 222$ km$^2$).

They have been chosen to cover two areas of the sea surrounding central Italy, close to the coastlines (as in previous studies [28,29]) and located along an ideal line passing through LNGS and perpendicular to the Apennine Mountains. In this work we will always refer to the sea waves height as the average of VHM0 data in the two marine domains.

In Sect. 5 we will show that the marine information extracted from the two reduced domains in Fig. 5b are a reliable approximation of the average status of the whole Adriatic and Tyrrhenian Seas (Fig. 5a), due to the fact that the typical size of Mediterranean storms is such to engulf large portions of the sea surrounding the Italian Peninsula. Therefore, the outcome of this work does not noticeably depend on the particular choice of the sea areas under investigation.

### 4.2 Correlation of CMEMS and seismometers data

Several measurements and models over decades assessed the power spectral density of the environmental noise and the sources of its different frequency components [15,16,30]. Many natural phenomena are sources of vibrations in the sub-Hz range. An example is represented by the marine microseismic activity, namely faint seisms caused by the motion of sea waves and by marine storms. They induce sub-Hz vibrations ($0.02 \lesssim \nu \lesssim 1$ Hz), which can propagate from the sea to the ground, and can therefore be detected also in the hinterland through seismometers. LNGS is located at $\simeq 50$ to $100$ km from the Italian coastlines, and it is therefore reached by microseismic waves coming from both the Adriatic and Tyrrhenian Seas.

In Sect. 3.2 we discussed the capability of identifying non-local seismic perturbations by correlating the responses of two distant seismometers (GIGS and SEISMO) at LNGS. This allows us to compare the time profile of the sea waves height provided by CMEMS with the response of any of the two seismometers. Figure 6 shows such comparison during September 2020. CMEMS detected the outbreak of a marine storm in the last ten days of the month. In the same days, GIGS and SEISMO detected an increase in the seismic activity at LNGS, with a time evolution identical to the storm development. Since any seismographic station at LNGS is affected by the combined microseismic waves from any direction from the Mediterranean Sea, we evaluate the Pearson correlation coefficients between GIGS data and the combined VHM0 data from both Adriatic and Tyrrhenian Seas. The correlation coefficient is $\rho_{GIGS,Seas} = 0.65$ in the first twenty days of September 2020, increasing up to $\rho_{GIGS,Seas} = 0.86$ in the last ten days of the month, during the storm development. Such high correlation between CMEMS and seismographic data highlights that the microseismic waves generated by the Mediterranean Sea activity can propagate up to LNGS and induce time-dependent perturbations, common to the entire LNGS and clearly detectable by seismometers.

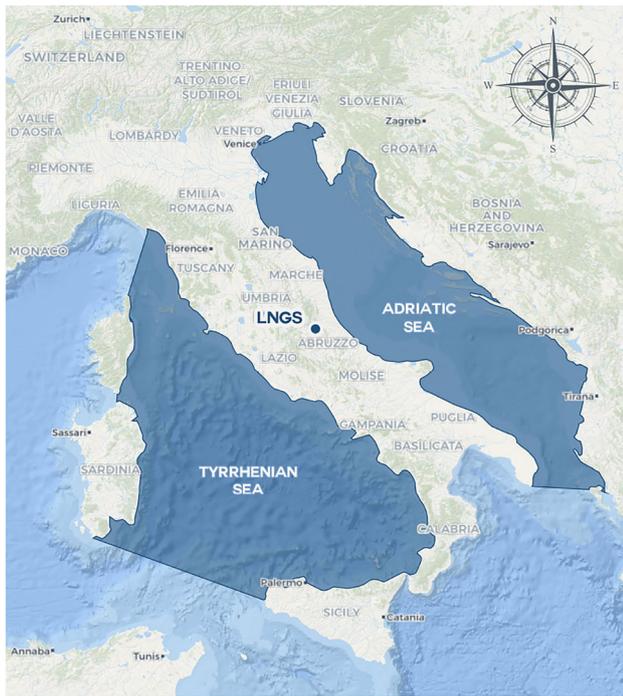

(a)

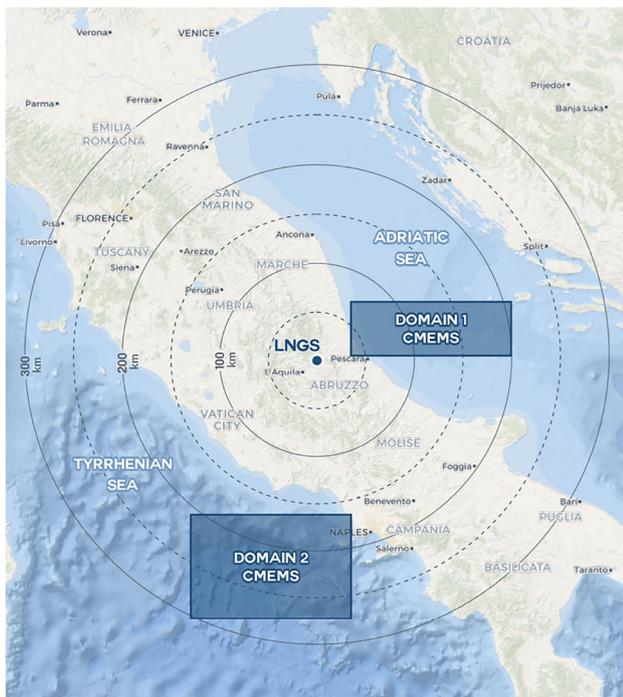

(b)

**Fig. 5** Map of the Adriatic and Tyrrhenian Seas (**a**) and of their respective domains extracted from CMEMS for the analysis in this work (**b**). The LNGS position and its distance from the two domains is also highlighted

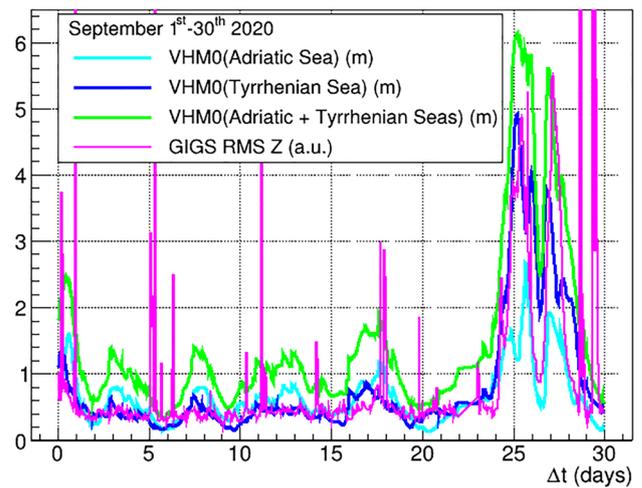

**Fig. 6** Comparison between the time profiles of VHM0 for Adriatic Seas, Tyrrhenian Sea and for their combined contribution, and the time profile of RMS Z of GIGS seismometer (rescaled by a constant factor for a better comparison with VHM0 data), during September 2020

We extended the study of the correlation between CMEMS and GIGS data to a period of more than three years, from January 1st, 2019 to July 31st, 2022 (Fig. 7). The time evolution of the correlation coefficients, evaluated month by month, shows a clear seasonal modulation, with maximum correlation during the boreal hemisphere winter and minimum during summer. Such modulation, with a period compatible with 1 year, is closely related to the modulation of the Mediterranean Sea activity. Indeed the sea activity is maximum during winter, being characterized by more intense and frequent storms and microseismic activity. Such microseismic noise is detected by seismometers at LNGS, resulting in a higher correlation between the seismic noise at LNGS and the status of the sea. Differently, the sea activity during summer is at its minimum and the induced microseismic noise contributed less to the total seismic noise at LNGS, which instead can be dominated by other sources of noise.

We established a chain of correlations between CMEMS data, providing insight on the Mediterranean Sea status, and seismometric data, measuring the induced microseismic activity at LNGS. Moreover, SEISMO seismometers share the same location of the CUORE experiment. Thanks to the established correlation between the various devices, we can directly correlate CMEMS and CUORE data, in order to determine if marine microseisms can affect the performance of CUORE low-temperature calorimeters.

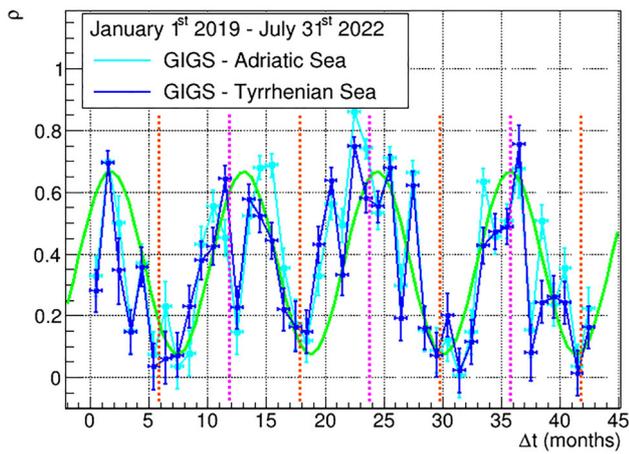

**Fig. 7** Time evolution of the correlation coefficient between the Adriatic and Tyrrhenian Seas activity (VHM0) with the GIGS response, from January 2019 to July 2022. The green line represents the outcome of a sinusoidal fit, whose period is compatible with 1 year. Dashed orange and pink lines represent respectively summer and winter solstices

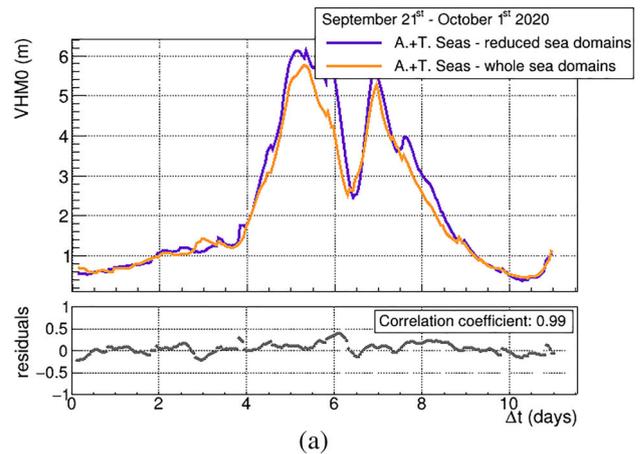

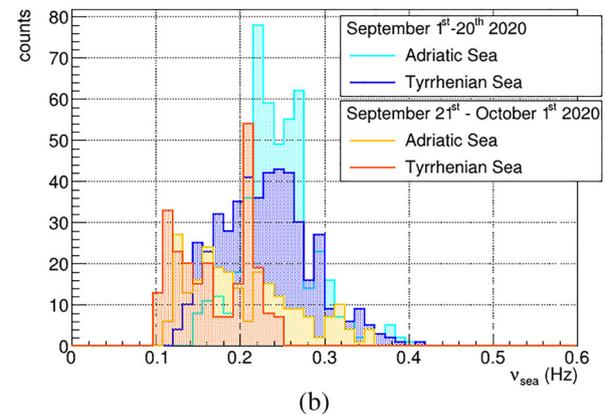

**Fig. 8** (**a**) Comparison between the time profiles of the sea waves height (combined contribution of Adriatic and Tyrrhenian Seas) evaluated, from September 21st to October 1st 2020, on the reduced sea domains in Fig. 5(b) and on the whole sea domains in Fig. 5(a); residuals and correlation coefficient between the two time profiles are also shown. (**b**) Sea waves frequency for Adriatic and Tyrrhenian Seas from September 1st to October 1st 2020

## 5 Correlation between marine microseisms and CUORE low-frequency noise

In this section we will assess and evaluate the impact of microseismic activity on the low-frequency noise of a pair of CUORE low-temperature calorimeters. Based on both CMEMS and seismometric data, we will focus on the time period from September 21st 2020 to October 1st 2020, during which a storm developed in the Mediterranean Sea. Figure 8a shows the corresponding time profiles of the sea waves height, for the combined contribution of Adriatic and Tyrrhenian Seas. It compares the sea waves height evaluated on the reduced sea domains in Fig. 5b and on the whole sea domains in Fig. 5a. The extremely high correlation between them proves that it is possible to reliably infer the status of the entire Adriatic and Tyrrhenian Seas from reduced sea domains, being the status of the sea dominated by the same environmental phoenomena.

Figure 8b shows the distribution of the sea waves frequency for Adriatic and Tyrrhenian Seas from September 1st to October 1st 2020. The characteristic frequency of sea waves is $0.1 \lesssim \nu_{sea} \lesssim 0.4$ Hz. During the stormy period, from September 21st to October 1st 2020, the sea waves frequency tends to be lower with respect to the previous period, characterized by a quieter sea.

### 5.1 Time evolution of CUORE low-frequency noise

The output voltage of CUORE detectors, referred to as baselines, shows the fluctuations and variations of their temperature, and is acquired and saved as a continuous stream, with a sampling frequency of 1 kHz. Subsequently, a software trigger is applied to such data streams to extract a random selection of noise fluctuations of the baselines, while a triggering algorithm is applied to identify thermal pulses. Time windows of a chosen length are then built around the noise/pulse triggers, defining noise events and pulse events.

In order to select a clean sample of noise-only events, we apply the following analysis cuts:

(a) reject time windows in which thermal pulses are triggered;

(b) reject time windows in which the maximum value of the baseline exceeds by five times its own RMS, to reject both events developing on decay tails of previous signal pulses and untriggered events featuring signals with significant amplitude;
(c) reject time periods affected by instabilities of the detectors and by earthquakes.

The most intense earthquakes, inducing instabilities in the baseline of CUORE detectors, are removed during periodic data-quality checks. Less intense seismic events are identified from the report provided by INGV. We reject all seismic events with epicenter within $\simeq 50$ km from LNGS with local magnitude LM$\geq$ 1.0, all seismic events with epicenter in Italy and with local magnitude LM$\geq$ 1.5, and all seismic events with epicenter abroad of Italy (teleseisms) and detected by the INGV monitoring system. For each of them, we reject a time period of ten minutes centered at the detection time provided by INGV. The rejected time window is longer in the case of teleseisms or earthquake swarms.

The frequency content of the CUORE detectors noise can be studied by applying a fast Fourier transform (FFT) routine to noise-only events. Marine microseisms are characterized by frequencies in the sub-Hz domain (Fig. 8b). The FFT sensitivity to low-frequency components of the noise can be enhanced by defining wide time windows for the detector events; for this analysis, we set the windows length to 60 s. By applying a FFT routine on noise data over a given time period we can build, for each detector, an average noise power spectrum (ANPS), representing the distribution of noise in the frequency domain. Given the characteristic frequencies of the marine phoenomena under investigation (Fig. 8b), we focus on noise components with frequencies below 1.4 Hz. In this way we can investigate the detectors response in a frequency interval wider than the one characterizing the source of marine microseisms, and assess if they react at the same frequency of the external microseismic-induced strain. This upper frequency limit at 1.4 Hz is set by a peak in the ANPS generated by the PT operation. The PT-induced noise is expected to be unaffected by the changes of the sea conditions, and therefore it is included in the analysis as a reference. Figure 9 shows the comparison of ANPS acquired by a CUORE detector during three different days: during quiet sea conditions, during a storm in the Mediterranean Sea, and during a high-intensity earthquake swarm with epicenter in central Italy. Several peaks appear at frequencies below 1 Hz. While their position is not affected by the storm development, their amplitude changes over time, increasing during the stormy day. Conversely, the amplitude of the PT-induced peak is almost unaffected. Figure 9 also shows that if time periods affected by high-intensity seismic events are not properly rejected, they can dominate the low-frequency noise, making it impossible to distinguish the effect of marine storms.

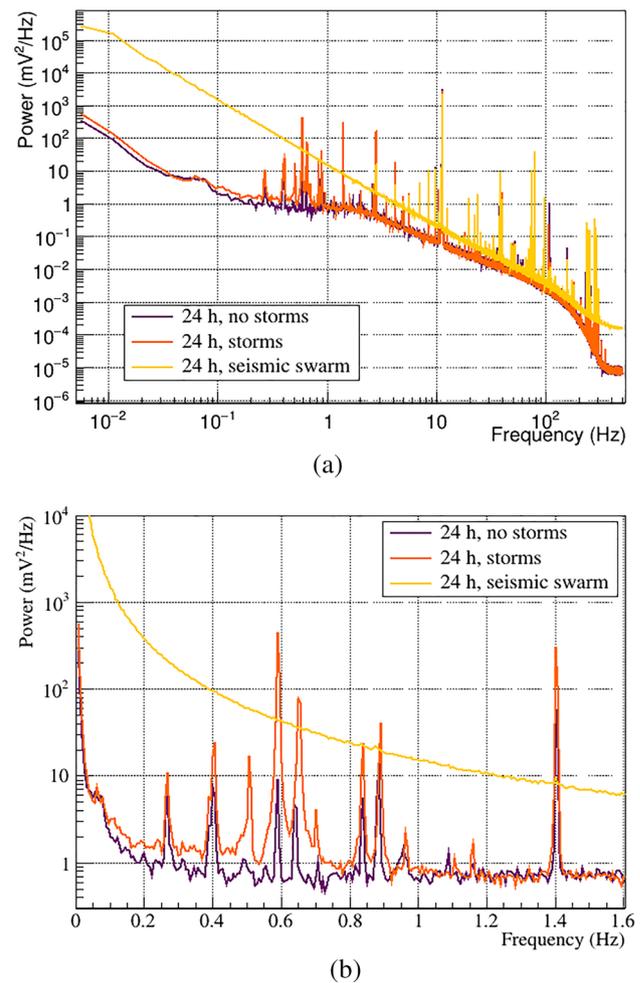

Fig. 9 Comparison of ANPS acquired by one CUORE low-temperature calorimeter in three different days: with and without storms in the Mediterranean Sea, and during a high-intensity earthquake swarm

In order to scan the evolution of the low-temperature calorimeters noise during the storm development, we generate an ANPS each $\simeq 12$ h. The integration of the ANPS in a given frequency range $\nu$ provides the power of the noise $P_\nu$ in that frequency interval. Since the position of the noise peaks is stable over time, we can define time-independent integration intervals (listed in Fig. 10).

The relative variation of $P_\nu$ over time, with respect to a reference period, provides a preliminary information of the sensitivity of the various frequency components of the noise to changes of the marine activity. Therefore we define the noise power ratio $R_\nu$:

$$R_\nu = \frac{P_{i,\nu}}{P_{ref,\nu}} \qquad (2)$$

where $P_{ref,\nu}$ is the noise power of a frequency component $\nu$ during a reference period of quiet sea activity and $P_{i,\nu}$ is the

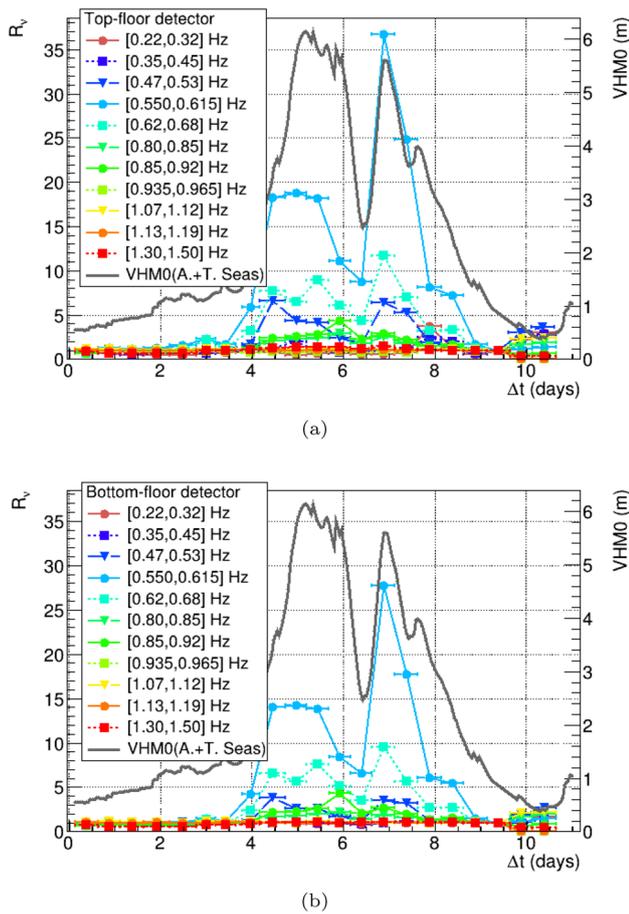

**Fig. 10** Comparison between the time evolutions of the sea waves height and of the noise power ratio $R_\nu$ for various frequency components of the noise, for two detectors placed at the top (**a**) and at the bottom (**b**) floors of a CUORE tower

noise power of the same frequency component in other time intervals $i$.

Figure 10 shows the comparison between the time evolution of the sea waves height (combining the contributions from Adriatic and Tyrrhenian Seas) and the time evolution of $R_\nu$ for various frequency components of the noise, for two detectors placed at the top (Fig. 10a) and at the bottom (Fig. 10b) a tower of CUORE low-temperature calorimeters. The comparison shows that the sub-Hz noise of CUORE detectors is excited when a storm develops in the Mediterranean Sea. However not all the frequency components are affected in the same way: noise at $\simeq 0.6$ Hz is the most excited, while noise above $\simeq 0.9$ Hz is almost unaffected.

### 5.2 Correlation between low-frequency noise and sea activity

To correlate the detectors low-frequency noise with marine microseisms we need to evaluate a proxy of the sea activity $I_S$, which we define as the integral over time of the sum of the wave height of both Adriatic Sea (VHM0$_A$) and Tyrrhenian Sea (VHM0$_T$):

$$I_S = \int_{t_i}^{t_f} [\text{VHM0}_A(t) + \text{VHM0}_T(t)] \, dt \qquad (3)$$

where $t_i$ and $t_f$ are the start and stop time of each time period over which the ANPS are generated. We associate to $I_S$ a systematic uncertainty due to the fact that the timing of CMEMS data may not coincide with the beginning/end of the 12 h long periods over which the detectors ANPS are built. We define such systematic uncertainty as the sea activity at the beginning and at the end of each time period over which the noise of CUORE detectors is evaluated.

Figure 11 shows that the power $P_\nu$ of various frequency components of the low-frequency noise of the two CUORE low-temperature calorimeters under study is linearly correlated with the sea activity $I_S$. By performing a linear fit for each frequency component of the noise, the angular coefficient $m_\nu^{rel}$, normalized by the minimum value of the corresponding noise power $P_\nu$, represents the relative variation of each noise component due to changes of the sea activity with respect to a period of quiet marine condition. Therefore, $m_\nu^{rel}$ quantifies the sensitivity of each noise component to changes of the sea activity.

Figure 12 shows the sensitivity profile of the low-frequency noise with respect to changes of the sea activity. Noise components at $\simeq 0.6$ Hz are the most sensitive to variations of the sea activity, while above $\simeq 0.9$ Hz the noise is almost unaffected to such environmental effects. The fact that CUORE detectors respond maximally at 0.6 Hz is a consequence of the peculiar structure of the CUORE suspension system, characterized by a resonant behaviour at $\simeq 0.5$ to 0.7 Hz. Microseismic waves, with characteristic frequencies mainly in the $0.1 \lesssim \nu_{sea} \lesssim 0.4$ Hz range (Fig. 8b), could excite such resonant mode, and therefore induce vibrations with corresponding frequencies to the detectors.

## 6 Conclusions

Here we demonstrate for the first time that low-temperature macro-calorimeters operated at $\simeq 10$ mK can be sensitive to environmental vibrations induced by marine microseisms, despite the sea being distant $\simeq 50$ to $100$ km from the detectors position. We presented a novel multi-detector approach to correlate environmental data about the Mediterranean Sea (provided by CMEMS), the seismic activity at LNGS (provided by GIGS seismometer from INGV and SEISMO from CUORE), and the response of a couple of CUORE low-temperature calorimeters. We assessed that marine microseisms induce an increase of the noise of CUORE detectors in the sub-Hz frequency range. Such noise is linearly correlated with the sea waves activity reconstructed by CMEMS

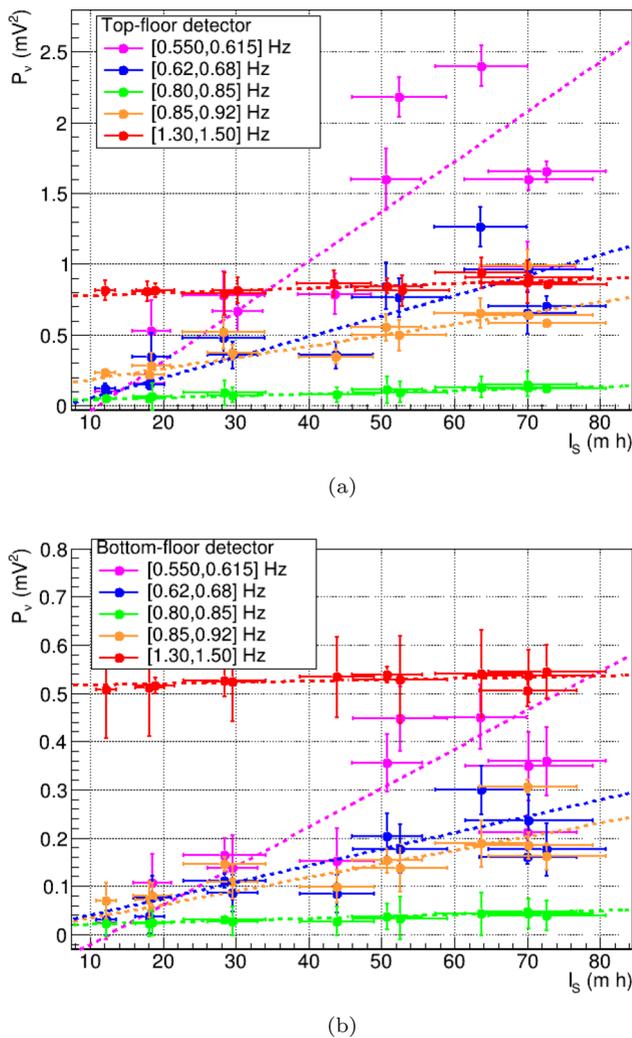

**Fig. 11** Noise power of five frequency components of the low-frequency noise as a function of the sea activity, for two detectors placed at the top (**a**) and at the bottom (**b**) floors of a CUORE tower

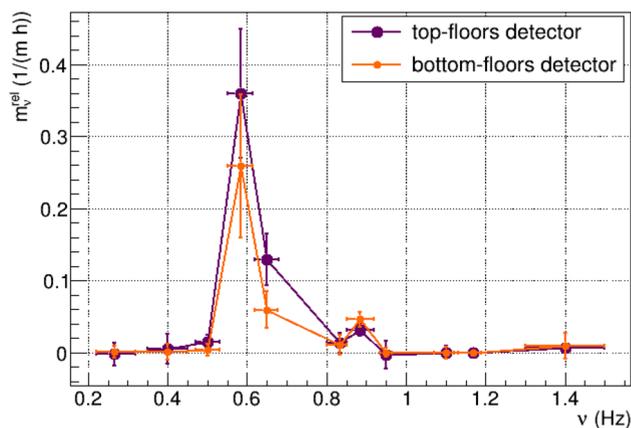

**Fig. 12** Sensitivity of the the low-frequency noise components to changes of the sea activity for two detectors, at the bottom and at the top of a CUORE tower

in two regions of the Adriatic and Tyrrhenian Seas close to the Italian coastline. The $\simeq 0.6$ Hz component of the CUORE noise is the most sensitive to variations of the marine activity, being the outcome of the excitation of a resonant mode of the CUORE suspension system by the marine microseism-induced vibrations.

The highlighted dependence between sub-Hz noise and sea activity can play a relevant role in limiting the energy resolution of macro-calorimeters, due to the impracticability of effectively filtering away such low-frequency noise, which belongs to the same frequency band of physics signals generated by energy depositions in the detectors.

The marine microseismic activity detected by the CUORE detectors is common to the entire LNGS underground facility, as supported by the combined detection by both GIGS and SEISMO seismometers. As a consequence such source of environmental noise can have relevant impact also on other cryogenic experiments hosted at LNGS which are searching for rare events, like $0\nu\beta\beta$ decay or dark matter interactions.

**Acknowledgements** This study has been conducted using E.U. Copernicus Marine Service information, and data from the GIGS seismic station from INGV. We thank the CUORE Collaboration for providing the data sample for the case study and for helping to make this work possible. This study was supported by Istituto Nazionale di Fisica Nucleare (INFN). The analysis reported in this work makes use of DIANA data analysis and APOLLO data acquisition software packages, which were developed by CUORICINO, CUORE, LUCIFER, and CUPID-0 Collaborations.

**Data Availability Statement** Data will be made available on reasonable request. [Authors' comment: The datasets generated during and/or analysed during the current study are available from the corresponding author on reasonable request.]

**Code Availability Statement** Code/software will be made available on reasonable request. [Authors' comment: The code/software generated during and/or analysed during the current study is available from the corresponding author on reasonable request.]